\documentstyle[aps,prl,twocolumn,psfig]{revtex}
\begin{document}
\twocolumn[\hsize\textwidth\columnwidth\hsize\csname
@twocolumnfalse\endcsname
\title{Equilibrium Distribution of Heavy Quarks in Fokker-Planck Dynamics}
\author{
\hfill D. Brian Walton$^a$ and Johann Rafelski$^b$
\hfill\raisebox{21mm}[0mm][0mm]{\makebox[0mm][r]{July 7, 1999}}%
}
\address{${}^a$Program in Applied Mathematics, and 
${}^b$Physics Department, University of Arizona, Tucson, AZ 85721\\
\hspace*{-1cm}E-Mail: walton@math.arizona.edu \hspace*{1cm}
 rafelski@physics.arizona.edu
}
\date{July 7, 1999}
\maketitle

\begin{abstract}\noindent
{We obtain within Fokker-Planck dynamics an explicit generalization of 
Einstein's relation between drag, diffusion and equilibrium distribution 
for a spatially  homogeneous system, considering both the
transverse and longitudinal diffusion for dimension \(n>1\).
We then provide a complete characterization of when the
equilibrium distribution becomes a Boltzmann/J\"uttner distribution, and when
it satisfies the more general Tsallis distribution. We apply this
analysis to recent calculations of drag and diffusion of a charm quark
in a thermal plasma, and show that only a Tsallis distribution 
describes the equilibrium distribution well. We also
provide a practical recipe applicable  to highly relativistic plasmas, 
for determining both diffusion coefficients so that a
specific equilibrium distribution will arise for a given drag
coefficient. \\

PACS number(s): 12.38.Mh, 05.10.Gg, 52.65.Ff, 05.20.Dd
}
\end{abstract}
\vskip0.5pc]
The velocity distribution of objects subject to a (thermal) 
background plays an
important role in a number of scientific fields, including
plasma physics and astrophysics \cite{Mel99}, 
nuclear physics \cite{Sve88,Mus98}, and, 
more generally, in kinetic theory\cite{Kam81,Ris89,Cse94}. 
The Fokker-Planck equation is a popular tool to study this
distribution. It can be motivated in a number of ways. One
method is to create a Langevin equation\cite{Kam81,Gar85}, which describes the
stochastic behavior of a single object propagating with random
noise. Another method comes by taking a master equation, such as the
linearized Boltzmann-Vlasov equation \cite{Cse94}, and performing a Landau
soft-scattering approximation. Finding recent theoretical 
calculations for the transport coefficients of the Fokker-Planck equation 
based on a microscopic theory \cite{Sve88,Mus98}, we recognized
the need to establish a simple procedure for understanding
the relation between the  transport coefficients of the Fokker-Planck
equations as determined in such microscopic calculations, and the 
resulting properties of the equilibrium  distribution. 
 
After a brief summary of the recent developments in the 
Fokker-Planck studies of equilibrating heavy quarks in 
a quark-gluon plasma,
we generalize and apply a well-known relation 
between kinetic coefficients and 
the equilibrium distribution of the Fokker-Planck 
equation in a spatially homogeneous environment \cite{LL1081}. 
In that way we are able to relate the drag and diffusion 
coefficients to the shape of the equilibrium distribution. 
We stress the importance of including both transverse 
and longitudinal diffusion to maintain a
consistent equation. A simple test
follows  which exactly determines when the
equilibrium distribution obeys
Boltzmann/J\"uttner statistics or the more
general Tsallis statistics \cite{Tsa88}. We also discuss 
how to choose the transport coefficients in order to attain the
Boltzmann/J\"uttner distribution, and address some issues related to 
the difference between the stopping power and the drag and diffusion 
coefficients.

The statistical properties of an ensemble of objects (particles)
 can be expressed in terms of
the one-particle distribution function, \(f(\vec{x},\vec{p},t)\). This
density, when multiplied by the \(2n\)-dimensional phase-space volume
element \(d^n\!{x}\, d^n\!{p}\), gives the probability of finding
the object in this infinitesimal region of phase-space. 
 We have introduced 
the dimensionality $n$ explicitly, and we will primarily pursue the 
physical case $n=3$, with the case $n=1$ also of interest due
to its exceptional character. We assume that \(f(\vec{x},\vec{p},t)\)
obeys a Boltzmann-Vlasov master equation of the form \cite{Cse94,LL1081}:
\begin{eqnarray}
  \frac{\partial}{\partial t} f \! + 
    {\dot {\vec x}} \cdot \!\nabla_x f  \!+
       {\dot {\vec p}} \cdot \!\nabla_p f  &= & \! \int  \!  \!
d^n{k} \, [W(\vec{p}+\vec{k},\vec{k})
                     f(\vec{x},\vec{p}+\vec{k},t)  \nonumber \\
&&\hspace*{0.8cm} -W(\vec{p},\vec{k}) f(\vec{x},\vec{p},t) ].
\label{FP1}
\label{BC1}
\end{eqnarray}
where:
\begin{equation}\label{FP_defs}
f=f(\vec{x}, \vec{p}, t),\quad
  {\dot {\vec x}}=\frac{d\vec{x}}{d t}=\frac{\vec p}{E},\quad
        {\dot {\vec p}}=\frac{d \vec{p}}{dt}=\vec F(\vec x)\,.
\end{equation}
In the non-relativistic limit, \(E\to m\), but otherwise our notation
is applicable to both classical and relativistic mechanics.  
The collision term has two parts: 
in the first gain term the transition rate \(W(\vec{p}_1,\vec{k})\) 
represents the rate that a particle with
momentum \(\vec{p}_1=\vec p+\vec k\) loses momentum \(\vec{k}\) due 
to reactions with the background. The second term represents 
loss due to scattering out. The collision term is strictly local, depending
only on momenta of particles, but it depends on position \(\vec x\) indirectly,
because \(W\) incorporates any background inhomogeneity.

Expanding the gain term  about \(\vec{p}\) to second order
in \(\vec{k}\) leads to the Fokker-Planck equation:
\begin{eqnarray}
\frac{\partial f}{\partial t} 
+ \dot x_i\frac{\partial f}{\partial x_i} 
+ \dot p_i\frac{\partial f}{\partial p_i} =
\frac{\partial}{\partial p_i} A_i f
+ \frac{\partial}{\partial p_i} \frac{\partial}{\partial p_j} B_{ij} f
\label{FP2}
\end{eqnarray}
where we are 
using the Einstein summation convention for repeated indices 
\(i\) and \(j\). We have introduced the transport coefficients 
of drag and diffusion, respectively:
\begin{eqnarray}
A_i=A_i(\vec{p}) &=& \!\int\! d^n{k} \, k_i \, W(\vec{p},\vec{k})\,,\nonumber\\
B_{ij}=B_{ij}(\vec{p}) &=&\frac12 \!\int\! d^n{k} \, k_i k_j \, W(\vec{p},\vec{k})\,.
\label{ABcof}
\end{eqnarray}

It is generally believed that the Fokker-Planck equation describes well 
the approach to (thermal) equilibrium. However, since we shall find that 
this is not guaranteed,   we record yet  another independent 
way to motivate the form of the Fokker-Planck Eq.\,(\ref{FP2}),
the Ito-Langevin method. Consider the Langevin system of equations:
\begin{equation}\label{lang}
\frac{d \vec x}{dt}=\frac{\vec p}{E}\,, \
\frac{d p_i}{dt}= F_i(\vec x) + G_i(\vec x,\vec p) + 
    D_{ij}(\vec x,\vec p\,) \eta_j(t) \,,
\end{equation}
where the noise term \(\vec \eta\) is Gaussian white noise with
 \(\langle \eta_i(t) \rangle = 0\), and 
\(\langle \eta_i(t) \eta_j(t') \rangle = \delta_{ij}\,\delta(t-t')\).
Using Ito's formula one shows that this Langevin system corresponds 
to the Fokker-Planck equation \cite{Gar85}:
\begin{equation}\label{lang_fp}
\frac{\partial f}{\partial t} + \frac{p_i}{E}\frac{\partial f}{\partial x_i}
+ F_i \frac{\partial f}{\partial p_i} =
\frac{- \partial G_i f}{\partial p_i} 
+ \frac{\partial^2 (\frac12 D D^T)_{ij} f}{\partial p_i \partial p_j}\,.
\end{equation}
Thus, we see that we can identify \(A_i \leftrightarrow - G_i\) and
\(B_{ij} \leftrightarrow (\frac12 D D^T)_{ij}\).

While the computation of \(D_{ij}\) is not obvious in the Langevin
formulation, the master equation approach gives precise formulas.
Written in terms of the two body collision reaction matrix elements
\(\cal M\), the drag and diffusion, according to   Eq.\,(\ref{ABcof}), 
are \cite{Sve88,Mus98}:
\begin{eqnarray}
A_i(\vec{p}) &=& \frac{1}{2 E_p}
\!\int\!\! {\frac{d^3 {k}}{(2\pi)^3 \, 2 E_k}}
\!\int\!\!  \frac{d^3 {k}^{\,'}}{(2\pi)^3 \, 2 E_{k'}}
\!\int\!\!  \frac{d^3 {p}^{\,'}}{(2\pi)^3 \, 2 E_{p'}} \times \nonumber \\
&&\hspace*{-1.5cm}\frac{1}{\gamma} \sum |{\cal M}|^2 (2\pi)^4
\delta^4(p+k-p'-k')[p_i-p'_i] g(\vec{k}) \tilde{g}(\vec{k}^{\,'}) \label{Acoll}\\
&\equiv& \langle\langle p_i-p'_i \rangle\rangle\,;\label{Adef}\\
B_{ij}(\vec{p}) &=&
\frac{1}{2}\langle \langle (p_i-p'_i)(p_j-p'_j) \rangle \rangle\label{Bdef}\,.
\end{eqnarray}
In our case, the incoming particle is different from the background.
For each background species, there is a similar 
additive contribution to  the collision integral in Eq.\,(\ref{Acoll}). 
Moreover, as long as the background does not distinguish
discrete quantum numbers of
incoming particles (such as spin), we can combine these properties in one
distribution \(f\). In such a case, we need to average over the initial
reaction channels and sum over the open final channels, akin to 
the evaluation of the cross section, hence the 
degeneracy factor \(\gamma^{-1}\) of the foreground particle. 
In Eq.\,(\ref{Acoll}) \(g(\vec{k})\) is the 
particle density of the background, assuming
that there is a single type of particle, and \(\tilde{g}
(\vec{k})=[1\pm g(\vec{k})]\) represents a Bose enhancement/Pauli
suppression factor for scattered background particles, as appropriate.
We assume that the background has equilibrated 
at the temperature \(T_{\rm b}\).

We are now prepared to study the equilibrium distribution and 
its relation with the drag and diffusion coefficients.
We study the simplest possible case
of a spatially homogeneous distribution. 
In absence of vectors other than \(\vec p\) the values of
\(A_i\) and \(B_{ij}\), which depend functionally on \(\vec{p}\)
and the background temperature \(T\), must be of the form, where 
\(p^2=p_i^2\) (summation convention is always implied):
\begin{eqnarray}
A_i(\vec{p},T)   &=& p_i A(p,T)\,,     \label{A_dep} \\ 
B_{ij}(\vec{p},T)&=& \left(\! \delta_{ij}\! 
   - \frac{p_i p_j}{p^2} \right) B_\bot(p,T) +
\frac{p_i p_j}{p^2} B_\|(p,T)\,.      \label{B_dep}
\end{eqnarray}
\(B_\|\) is the longitudinal and \(B_\bot\) the transverse diffusion
coefficient.  In terms of microscopic reaction amplitudes, these
three functions are defined by the following expressions \cite{Sve88}:
\begin{eqnarray}\label{A_def}
A(p) &=& \langle \langle 1 \rangle \rangle - \frac{\langle \langle \vec{p}
        \cdot \vec{p}{\,'} \rangle \rangle}{p^2}\,,\\
\label{B0_def}
B_\bot(p) &=& \frac{1}{4} \left[ \langle \langle p'^2 \rangle \rangle
 - \frac{\langle \langle (\vec{p} \cdot \vec{p}{\,'})^2 \rangle \rangle }
   {p^2}\right]\,,\\
B_\|(p) &=& \frac{1}{2} \left[
 \frac{ \langle \langle (\vec{p} \cdot \vec{p}{\,'})^2 \rangle \rangle }{ p^2}
 - 2 \langle \langle \vec{p} \cdot \vec{p}{\,'} \rangle \rangle
 +  p^2 \langle \langle 1 \rangle \rangle\right]. \hfill
\label{B_def}
\end{eqnarray}

With no external forces and a homogeneous background, 
Eq.\,(\ref{FP2}) reads:
\begin{equation}
\frac{\partial f}{\partial t}
= \frac{\partial}{\partial p_i} \left(A_i f
+ \frac{\partial}{\partial p_j} B_{ij}f\right)=
 -\vec \nabla_p \cdot\vec {\cal P}\,.
\label{eq:Fokk_Planck_Genl}
\end{equation}
A natural requirement for \(f_{\rm eq}\), detailed balance, is for the
probability current \(\vec{\cal P}\) to  vanish  \cite{LL1081}:
\begin{equation}
A_i(\vec{p},T) = B_{ij}(\vec{p},T) 
\frac{\partial \Phi(\vec p)}{\partial p_j}
 - \frac{\partial B_{ij}(\vec{p},T)}{\partial p_j}\,,
 \label{pot_cond}
\end{equation}
where we wrote  the equilibrium distribution as:
\begin{equation}
f_{\rm eq}( p;T,q) = N \exp(-\Phi(p;T,q))\,.
\label{SS}
\end{equation}
where $T,q$ are parameters that may be needed to characterize
the shape of the distribution. 
R. Graham and H. Haken \cite{Gra71ab,Gra80} provide a more general
approach which classifies when such a simplification is valid, and
how to extend the condition Eq.\,(\ref{pot_cond}) when it is not valid. 
For our case,  we have verified that Eq. (\ref{pot_cond}) is valid.

Using Eqs.\,(\ref{A_dep},\ref{B_dep}) and the fact that in the
spatially homogeneous case the equilibrium distribution depends only on
\(p=|\vec{p}\,|\),  Eq.\,(\ref{pot_cond}) becomes
\begin{eqnarray}
A(p,T) &=& \frac{1}{p} \frac{d \Phi}{d p} B_\|(p,T)
                    - \frac{1}{p} \frac{dB_\|}{dp}\nonumber\\
&&- \frac{n-1}{p^2} (B_\|(p,T)-B_\bot(p,T))\,, \label{GenEin}
\end{eqnarray}
which is the desired  relation between the shape of equilibrium
distribution and the three drag/diffusion coefficients.
The special case considered by Einstein  arises in the 
classical problem of a particle that
travels through an ideal heat bath and undergoes linear 
damping (Rayleigh's particle). Substituting the
coefficients \(A = \gamma\) and \(B_\bot=B_\|=D\) into the relation
(\ref{GenEin}), we obtain the Boltzmann equilibrium
distribution Eq.\,(\ref{SS}) with  \(\Phi = p^2/2mkT\)
only if Einstein's well-known drag-diffusion relation for Brownian 
motion \(\gamma = D/mkT\) is satisfied.

When the equilibrium distribution is known {\it a
priori} and if the diffusion coefficients are also known, then it is an easy
matter to use Eq.\,(\ref{GenEin})
to find the unique, consistent drag coefficient. However, the reverse is not 
 true for \(n>1\): given the equilibrium distribution and the 
drag coefficient, there are two diffusion coefficients which must be
simultaneously determined, which is in general not possible without a
further assumption.  One `popular' method to
do this is to assume that the tensor \(B_{ij}\) is diagonal, that is 
\(B_\bot=B_\|\) and then to solve the linear first order
differential equation  to obtain \(B_\|\):
\begin{equation}
\label{recA}
\frac{d}{dp}\left( e^{-\Phi(p)} B_\|(p) \right) + p A(p) e^{-\Phi(p)} = 0\,.
\end{equation}
An alternative method, motivated when \(T_{\rm b}\) is the only dominant energy
scale, is to calculate the longitudinal diffusion term \(B_\|\) in terms
of the drag and then to determine the transverse diffusion term \(B_\bot\) :
\begin{equation}
B_\| \to \frac{p A}{d\Phi/dp}\,,\qquad
\label{rec_step1}\label{rec_step2}
B_\bot \to B_\| + \frac{p}{n-1}\frac{dB_\|}{dp}\,.
\end{equation}
We note that when \(B_\|=\mbox{const.}\) both recipes
give the same result. 
These approximate relationships are sufficient to
guarantee correct equilibrium behavior, but may not
represent accurately the dynamical evolution, and 
one should always require non-negative diffusion.

While discussing the relationships between the drag and diffusion coefficients
of the Fokker-Planck equation, we note another interesting relation
between these coefficients and the frequently discussed stopping power.
The stopping power measures the energy loss per unit distance traveled,
and is equal to the 
energy loss per unit time divided by the particle speed.
Thus,  in terms of the elementary matrix elements \cite{Bra91a}:
\begin{equation}
- \frac{dE}{dx} = \langle \langle \frac{E-E'}{v} \rangle \rangle=
\langle \langle \frac{E^2-E\,E'}{p} \rangle \rangle\,.
\end{equation}
When combined appropriately with \(A_i\), a relativistically 
invariant (scalar) quantity is found:
\begin{eqnarray}
 p_i A_i + p\frac{dE}{dx} &=& 
 \langle \langle 
    \vec{p} \cdot (\vec{p} - \vec{p}{\,'})-E^2 +E E' 
 \rangle \rangle \label{3rdTr}\\
&=& -\frac12\langle \langle (p_\mu -\, p'^\mu)^2 \rangle \rangle\,,
\nonumber 
\end{eqnarray}
where we use four-vector notation. Eq.\,(\ref{3rdTr}) shows  that the 
stopping power and the drag coefficient are, in general,  two 
independent quantities. To connect them we need to evaluate also:
\begin{equation}
B_{00}\equiv \langle \langle (E-E')^2\rangle \rangle\,,\quad
A_0\equiv \langle \langle E-E'\rangle \rangle\,.
\end{equation}
In the non-relativistic limit these two new quantities are relatively small.
The energy loss can be expressed as:
\begin{equation}
-\frac{dE}{dx} = \frac{p_i A_i +B_{00}- B_{ii}}{p}\,,\qquad 
-\frac{dE}{dt} = A_0\,.
\label{eloss}
\end{equation}
For the Rayleigh particle considered earlier, 
noting \(B_{00}\to 0\), this corresponds to an energy loss:
\begin{equation}
-\frac{dE}{dx} = \gamma \frac{2m}{p}
       \left( \frac{p^2}{2m} - \frac{3kT}{2} \right)=
                 \gamma m (v-\frac{\langle v^2 \rangle_T}{v})\,,
\label{eloss_ray}
\end{equation}
which vanishes precisely for a thermal velocity.

The problem that we are facing even in the simple spatially homogeneous
case is that the Fokker-Planck coefficients cannot simply be chosen
to assure that the `correct' equilibrium distribution results but 
have already been obtained in terms of elementary collision 
reaction amplitudes, see Eqs.\,(\ref{A_def}--\ref{B_def}), 
 and thus the resulting equilibrium distribution
is fixed, as can be seen solving and integrating 
Eq.\,(\ref{GenEin}) to obtain \(\Phi\).  Since the drag and diffusion
coefficients are not evaluated exactly but in some valid approximation,
typically applying a perturbative expansion, it is more appropriate to
analyze the resulting distribution in terms of some useful class.
We consider the class of Tsallis statistics \cite{Tsa88}, which depends
on a temperature-like quantity  \(T\) and on a parameter  \(q\),
which measures the degree of extensivity of entropy in the system:
\begin{eqnarray}\label{TsDis}
f_{\rm eq} &=& N \left[ 1 - (1-q) E(p)/T \right]^\frac{1}{1-q}\,,\\
\frac{d\Phi}{dp} &=& \frac{dE}{dp}\frac1{T - (1-q)E}\,.\label{TsCon}
\end{eqnarray}
The Boltzmann distribution arises when \(q \to 1\).  
Substituting into Eq.\,(\ref{GenEin}), we obtain:
\begin{equation}
T + (q-1) E = \frac{dE}{dp}\frac{ B_\|}{p A + \frac{dB_\|}{dp}
        + \frac{n-1}{p}(B_\|-B_\bot)}.
\label{t_fit}
\end{equation}
Whenever the ratio given by the right-hand side 
of Eq.\,(\ref{t_fit}) becomes linear
in \(E\), then Tsallis statistics describe the 
stationary distribution. When the
ratio is constant, then a Boltzmann/J\"uttner 
distribution suffices. 
We note that for the special case \(n=1\) and non-relativistic dynamics
\(dE/dp=v\), Eq.\,(\ref{t_fit}) was obtained recently within the 
Langevin formulation of the Fokker-Planck dynamics \cite{Bor98}.

\begin{figure}[htb]
\centerline{\psfig{figure=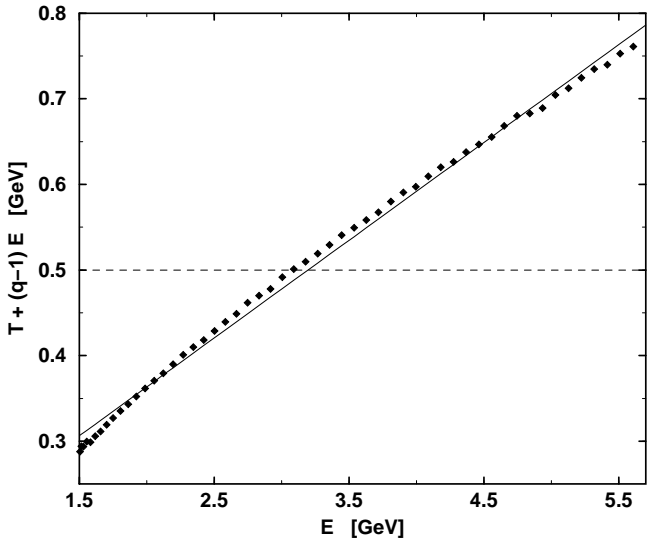,width=3.375in}}
\caption{Calculated data (diamonds) and linear fit 
for the ratio in Eq. (\ref{t_fit}) for a charmed quark \(m_c=1.5\)\,GeV
thermalizing in gluon background at \(T_{\rm b}=500\)\,MeV. 
Dashed line: result expected for a 
Boltzmann/J\"uttner distribution, \(T=T_{\rm b}\).}
\label{fig:ratio}
\end{figure}

We consider now the drag and diffusion
coefficients for a charm quark with mass \(m_c=1.5\)\,GeV  
interacting with thermal gluons at \(T_{\rm b}=500\)\,MeV 
calculated using perturbative QCD techniques \cite{Sve88,Mus98}. 
We have gone to
considerable length to assure that these results apply \cite{Sri98}. 
Diamonds in Fig.\,\ref{fig:ratio}  show the ratio  Eq.\,(\ref{t_fit}). 
The linear regression fit (straight line) shows that
the parameters best describing the distribution as a Tsallis distribution
are \(q=1.114\) and \(T_{\rm T}=135.2 \)\,MeV. 
The dashed horizontal line in Fig.\,\ref{fig:ratio} corresponds to the 
Boltzmann/J\"uttner distribution (\(q=1)\) and \(T_{\rm T}=T_{\rm b}\)), which
we were expecting to find. The difference to the 
actual distribution appears to be significant.

\begin{figure}[htb]
\centerline{\psfig{figure=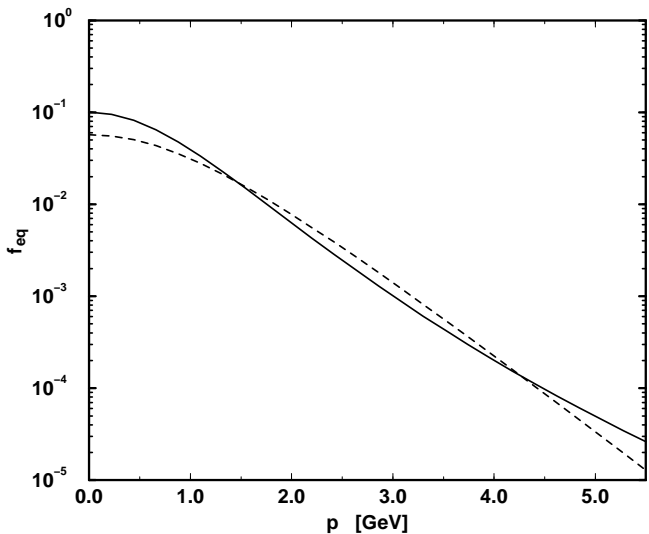,width=3.375in}}
\caption{Normalized equilibrium spectra:
 Tsallis distribution (solid line) 
and   Boltzmann/J\"uttner distribution (dashed line)
at \(T_{\rm b}=500\)\,MeV, with  \(m_c=1.5\) GeV.}
\label{fig:spec}
\end{figure}

The more practical question is what the charmed quark
spectrum would actually look like. This is shown in Fig.\,\ref{fig:spec}
where a solid line shows the Tsallis distribution as obtained above, 
compared to Boltzmann/J\"uttner shape  for  
\(T_{\rm b}=500\)\,MeV and \(m_c=1.5\) GeV.
Assuming that  the Tsallis shape would be measured, the
spectrum would reveal two components: at low \(E\)  a  `cold'
Boltzmann distribution, and for  high \(E\)  a power-law  with 
\(f_{\rm eq}\propto E^n\), where in our case \(n=\frac{1}{1-q}=-8.8\).

In summary, we 
developed tools which allow one to
identify within Fokker-Planck dynamics
the equilibrium distribution for given (calculated) 
drag and diffusion coefficients, or when the stationary distribution
is known, to determine the drag or, as a recipe for \(n>1\) dimensions, 
both longitudinal and transverse diffusion coefficients. 
We have then shown that thermalization of charmed quarks 
in a quark-gluon plasma leads to the two parameter 
Tsallis  distribution, and have determined the pertinent parameters 
for the published microscopic drag/diffusion coefficients.
It is at present unclear if the resilience of the Tsallis 
statistics is a fundamental feature of the Fokker-Planck 
dynamics in background relativistic plasma \cite{Kan96}.
It is important to note that only a major change in the transport 
coefficients from the results of the microscopic calculations
will lead to a Boltzmann/J\"uttner equilibrium distribution.

{\it Acknowledgments:\/}
We thank D. Pal and D.K. Srivastava for helpful correspondence,
E.D. Davis for valuable comments and suggestions,
T. Kodama for introducing us to Tsallis statistics, and 
G. Eyink for pointing out to us the work by Graham and Haken.
Work supported in part by a grant from the U.S. 
Department of Energy,  DE-FG03-95ER40937\,. B. Walton:
This material is based partially  upon work 
supported under a National Science
Foundation Graduate Fellowship.

\end{document}